\documentclass[twocolumn,twoside,slac_two]{revtex4}
\usepackage{graphicx}
\usepackage{fancyhdr}
\newcommand{\etal}{\textit{et al.}}

\begin{document}

\title{Beyond the CDMS-II Dark Matter Search: SuperCDMS}

%

\author{P.L. Brink, B. Cabrera, C.L. Chang, J. Cooley, R.W. Ogburn}
\affiliation{Department of Physics, Stanford University, Stanford, CA 94305, USA}

\author{D.S. Akerib, C.N. Bailey, P.P. Brusov, M.R. Dragowsky, D.R. Grant, 
R. Hennings-Yeomans, R.W.~Schnee}
\affiliation{Department of Physics, Case Western Reserve University, 
Cleveland, OH 44106, USA}

\author{M.J. Attisha, R.J. Gaitskell, J-P.F. Thompson}
\affiliation{Department of Physics, Brown University, Providence, RI 02912, 
USA}

\author{L. Baudis, T. Saab}
\affiliation{Department of Physics, University of Florida, Gainesville, FL 
32611, USA}

\author{D.A. Bauer, M.B. Crisler, D. Holmgren, E. Ramberg, J. Yoo}
\affiliation{Fermi National Accelerator Laboratory, Batavia, IL 60510, USA}

\author{R. Bunker, D.O. Caldwell, R. Mahapatra, H. Nelson, J. Sander, S. 
Yellin}
\affiliation{Department of Physics, University of California, Santa 
Barbara, CA 93106, USA}

\author{P. Cushman, L. Duong, A. Reisetter}
\affiliation{School of Physics \& Astronomy, University of Minnesota, 
Minneapolis, MN 55455, USA} 

\author{P. Denes, A. Lu, B. Sadoulet}
\affiliation{Lawrence Berkeley National Laboratory, Berkeley, CA 94720, USA}

\author{J. Filippini, P. Meunier, N. Mirabolfathi, B. Sadoulet, D.N. 
Seitz, B. Serfass, K.M. Sundqvist}
\affiliation{Department of Physics, University of California, Berkeley, CA 
94720, USA}

\author{S.R. Golwala}
\affiliation{California Institute of Technology, Pasadena, CA 91125, USA}

\author{M.E. Huber}
\affiliation{Department of Physics, University of Colorado at Denver and 
Health Sciences Center, Denver, CO 80217, USA}

\author{K.D. Irwin}
\affiliation{National Institute of Standards and Technology, Boulder, CO 
80303, USA}

\author{B.A. Young}
\affiliation{Department of Physics, Santa Clara University, Santa Clara, 
CA 95053, USA}

\begin{abstract}
Presently the CDMS-II collaboration's Weakly Interacting Massive Particle (WIMP) 
search at the Soudan Underground Laboratory sets the most stringent exclusion 
limits of any WIMP cold dark matter direct-detection experiment. To extend our reach 
further, to WIMP-nucleon cross sections in the range $10^{-46} - 
10^{-44}$cm$^2$, we propose SuperCDMS, which would take advantage of a 
very deep site. One promising site is the recently approved SNOLab facility 
in Canada. In this paper we will present our overall plan, identify primary 
issues, and set the goals that need to be met prior to embarking upon each phase 
of SuperCDMS.

\end{abstract}

\maketitle

\thispagestyle{fancy}


\section{INTRODUCTION}

%


The identification of dark matter
is of fundamental importance to cosmology, astrophysics and high-energy particle 
physics~\cite{Bergstrom}.
Over the last decade a variety of cosmological observations, from the primordial abundance
of light elements to the study of large-scale structure, in combination with high-redshift
supernovae findings, weak-lensing observations, and detailed mapping of the anisotropy of
the cosmic microwave background, have led to the construction of a concordance model of
cosmology. In this very successful model, the universe is made of $\sim$ 4\% baryons which
constitute the ordinary matter, $\sim$ 23\% non-baryonic dark matter which dominates structure
formation, and $\sim$ 73 \% dark energy~\cite{Spergel}.

A leading hypothesis is that the dark matter is comprised of Weakly Interacting Massive
Particles, or WIMPs, a hypothetical elementary particle that was produced moments after
the Big Bang. Supersymmetry extensions to the standard model of particle 
physics predict a stable particle with the appropriate properties, the Lightest 
Supersymmetric Particle (LSP)~\cite{Jungman}.

If WIMPs are indeed the dark matter they can be detected via elastic 
scattering from nuclei in a suitable target~\cite{Primack}. 
The predicted energy depositions and event rate of these expected nuclear 
recoils are low. 
The direct WIMP-search experiments, in 
particular CDMS~II~\cite{PRL2004}, EDELWEISS~\cite{EDELWEISS}, and 
ZEPLIN~I~\cite{ZEPLIN} are beginning to set significant upper limits on 
the WIMP-nucleon scattering cross-section.
The CDMS ZIP (Z-dependent Ionization and Phonon) detectors, which are 
sensitive to both ionization and athermal phonon signals,
have demonstrated extraordinary selectivity for the nuclear recoils expected 
from WIMP interactions and are currently the most sensitive detectors in the search 
for WIMPs~\cite{PRL2004}.

While the present direct WIMP-search experiments, with target masses of a few kgs,
already provide interesting constraints on the hypothetical WIMP particle properties
(and thus constraints on models of supersymmetry), there is a strong demand to perform 
more sensitive direct WIMP searches with larger detector mass.

\begin{figure}[thbp]
\centering
\includegraphics[width=75mm]{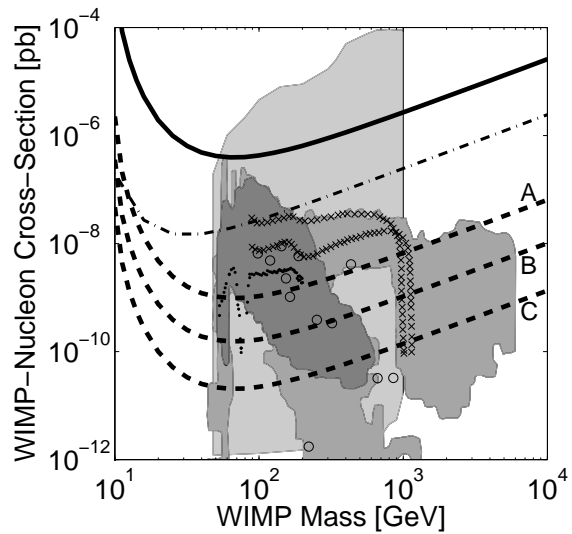}
\caption{Present CDMS~II (solid curve)~\cite{PRL2004} WIMP exclusion limit 
(90\% C.L. above curve); projected sensitivity goal of CDMS~II (dot-dashed curve), and projected 
SuperCDMS phases A, B \& C (labelled dashed curves). The lightest grey region is 
the allowed WIMP-nucleon cross-section under standard model
assumptions for the Minimal SuperSymmetric Model (MSSM) parameter 
space~\cite{Kim}. SuperCDMS will probe nearly all split-supersymmetry 
models ($\times$'s~\cite{Giudice} and dots~\cite{Pierce}) and much of the 
mSUGRA region~\cite{Baltz} (medium grey), including most post-LEP benchmark 
points (circles)~\cite{Battaglia} and nearly all the subset (dark grey) 
consistent with a supersymmetric interpretation of the muon $g-2$ 
measurement.
} \label{fig:Goal}
\end{figure}

Figure~\ref{fig:Goal} shows the present CDMS~II 
WIMP-exclusion limit and the allowed regions of mass and cross-section that arise 
from a number of favoured supersymmetry models. Also shown is 
the SuperCDMS program which is proposed to increase the ultimate WIMP-search 
reach by a factor of a few hundred past that of CDMS-II. This would allow 
a sensitivity in the WIMP-nucleon scalar cross-section to complement that 
of upcoming supersymmetry searches, notably at the Large Hadron Collider (LHC).

Figure~\ref{fig:timeline} shows the 
overall time-line, including the commissioning and data-taking for each 
proposed SuperCDMS phase.
Note that the detector deployment 
is phased, commensurate with the fabrication and testing schedule required 
for the new ZIP-style detectors.  

Funding approval for SuperCDMS would enable a new generation of direct 
detection experiments utilizing ZIP-style detectors with target masses 
starting at 27~kg (Phase~A), growing to 145~kg (Phase~B), and 
up to 1100~kg (Phase~C) to study a potential WIMP signal.

\section{SUPERCDMS}

The goal of each phase of SuperCDMS is to operate with a nuclear-recoil 
event background close to zero. The resultant WIMP-search sensitivity 
would then scale linearly with exposure and allow each phase of SuperCDMS 
to be conducted within a few years. 

Scaling from the recent CDMS Soudan 
results~\cite{PRL2004},
the sensitivity goal of $2\times 10^{-45} {\rm cm}^2$ for SuperCDMS Phase~A 
(for a 60~GeV mass WIMP, see Fig.~\ref{fig:Goal}) corresponds to an event rate of
$4 \times 10^{-4}{\rm /kg/day}$, integrated over the nuclear recoil energy 
range of interest: 15-45~keV. However, this sensitivity goal is for 
setting a 90\% confidence WIMP-exclusion upper limit. The desired ``zero-background'' 
contamination level required is $ \sim 1 \times 10^{-4}{\rm /kg/day}$ for 
15-45~keV. Experience with the ZIP detectors of CDMS suggests an integrated 
exposure of one to two years with 27~kg of Ge ZIP-style detectors
would be required to reach the SuperCDMS Phase~A goal, assuming zero 
background events.

\subsection{Site Selection: Neutron Background}

The CDMS~II WIMP-search experiment is operated at the Soudan Mine with an 
overburden of 2090~meter-water-equivalent (m.w.e.) and an active muon-veto shield and substantial
polyethylene, all to mitigate neutron backgrounds~\cite{PRL2004}. 
The neutron background at Soudan that could mimic 
WIMP events in the CDMS~II experiment is estimated to be
$4 \times 10^{-4}{\rm /kg/day}$ (with an uncertainty of a factor of 
two) for the Ge nuclear recoil energy range of interest, 
15-45~keV~\cite{PRD2005}. Thus, the 
SuperCDMS Phase~A zero-background goal cannot be accomplished at the Soudan 
site with our present shielding configuration. Although further shielding
measures at Soudan could accomplish the Phase~A goal, later SuperCDMS phases 
would still remain out of reach. 

During the last year
a significant development occurred with the 
approval in Canada of the SNOLab deep-site facility at the Sudbury Mine. 
The $\sim 6000$~m.w.e. overburden at this 
site results in over two orders of magnitude 
suppression in the neutron background compared to Soudan and would even allow 
SuperCDMS~Phase~C (with 1100~kg of Ge detectors running for several years) 
to be accomplished. The SuperCDMS proposal has generated strong interest at 
SNOLab and further discussions are in progress.

\begin{figure*}[htbp]
\centering
\includegraphics[width=110mm]{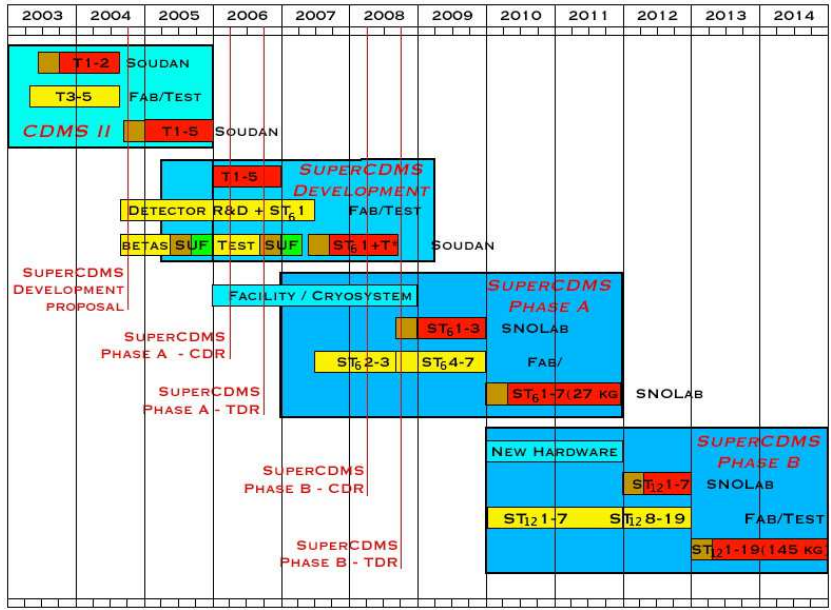}
\caption{Proposed timeline for the SuperCDMS Development program and 
SuperCDMS Phases A \& B. Note the phased deployment of detectors to allow 
early science returns, even in the Development program.
The WIMP-search running of the 27~kg of Ge of Phase A and the 145~kg of Ge of Phase B 
are proposed to occur at SNOLab.
} \label{fig:timeline}
\end{figure*}

With the possibility of SNOLab occupancy in 2007, Fig.~\ref{fig:timeline} 
shows a possible timeline for the execution of the SuperCDMS program. 
The Phase~A WIMP-search would commence at the new deep site in a new cryogenic 
system, which would also be used for the following Phases B \& C. 

\subsection{Detector Performance: Gamma and Beta Backgrounds}

Recent results from the CDMS~II detectors at Soudan~\cite{PRL2004,PRD2005} confirm the excellent 
discrimination ability of the ZIP detectors against the most common source 
of background events, photons from external sources of radioactivity 
interacting in the bulk of the detector. 
Thus, the 
expected gamma background of SuperCDMS ($\sim 1$/keV/kg/day) is not 
an immediate source of concern for SuperCDMS Phases A~\&~B. 
However, a more problematic source of backgrounds that does need to be 
addressed prior to SuperCDMS Phase~A is that due to electromagnetic events 
near the detector surfaces.

Electromagnetic 
interactions within the first $35\,\mu{\rm m}$ of the present ZIP detector 
surface give a suppressed ionization signal, and for events within the 
first $1\,\mu{\rm m}$ of the surface the suppression is sufficiently severe 
that the ionization yield measured mimics nuclear recoils.
The great advantage of 
the ZIP detectors of CDMS is that the athermal phonon signal for such 
surface events have measurably faster risetimes. 
Thus software analysis cuts can be established to remove them.

The CDMS~II data analysis approach for each 
WIMP-search run at Soudan has been to use the {\it in-situ} photon and 
neutron calibrations, laboratory calibrations with beta sources, and their 
Monte-Carlo simulations
to establish event identification cuts, which are estimated to leave less than 
one ``beta-leakage'' event in the (blinded) nuclear-recoil WIMP-search data~\cite{PRL2004,PRD2005}. 
Clearly, the more aggressive the cuts need to be to reduce the beta-leakage 
estimate, the lower the net efficiency for 
true nuclear recoil identification and the 
less sensitive the WIMP-search becomes.
No statistical subtraction of beta events in the WIMP-search data 
is presently attempted, nor assumed for SuperCDMS. 

Studies~\cite{PRD2005} of the detector stacking configuration used by CDMS 
show that single-scatter events caused by electrons ejected 
from nearby material by an incident photon have a depth distribution in 
the detector that is deeper, and thus 
less problematic, than electrons of similar deposited energy emitted from 
beta sources on the detector surfaces. With the expected gamma background of SuperCDMS, the 
ejected-electrons due to photon-induced events are not a source of concern. 
Instead, the source of 
background that needs to be reduced is electrons due to beta
contamination. Here we take beta contamination to include beta-emitters, 
internal-conversion sources, Auger electrons and soft X-ray sources as well, 
although the source is most likely to be a beta-emitter.

The present ZIP detectors' discrimination performance and data analysis
results in an inferred ``mis-identified beta'' background due to beta-emitter 
contaminants of $ \sim 1 \times 10^{-2}{\rm /kg/day}$ for the 
15-45~keV nuclear recoil energy range of interest~\cite{PRL2004,PRD2005}.
This is a factor of one hundred too high for Phase~A. 

\subsection{Detector Advances Required}

In order to achieve the sensitivity goals of the proposed SuperCDMS Phases A \& 
B (with zero background events), a number of improvements are required to the 
present CDMS II ZIP detector technology to reduce the inferred 
mis-identified beta background. We have identified three approaches, all of 
which we believe will contribute at a similar level: 
improved ZIP-style detector performance, advances in analysis techniques, 
and actual reduction of the source of beta-contamination itself.
Table~\ref{tab:one} summarizes our expectations for 
the target improvement factors that will contribute to both Phases A~\&~B. 

\begin{table}[tbhp]
\begin{center}
\caption{Target improvement factors to achieve the SuperCDMS Phase A and 
Phase B WIMP-search sensitivities. The first and third columns are relative to the 
present CDMS~II ZIP detector performance at Soudan~\cite{PRL2004}.
}
\begin{tabular}{|l|c|c|c|c|}
\hline \textbf{Improve} & & \textbf{Phase A} & \textbf{Phase B} & \textbf{Combined} \\
\hline Detector & &  &  &  \\
 rejection & & $\times 5$ & $\times 2$ & $\times 10$ \\
\hline Analysis & & &  & \\
 discrimination & & $\times 4$ & $\times 2$ & $\times 8$ \\
\hline Contamination & & & & \\
 reduction & & $\times 5$ & $\times 2$ & $\times 10$ \\
\hline
\hline \textbf{Total} &  & &  & \\
 \textbf{improvement} & & $=\times 100$ & $=\times 8$ & $=\times 800$ \\ 
\hline
\end{tabular}
\label{tab:one}
\end{center}
\end{table}

Referring to Table~\ref{tab:one} and focussing on the requirements for 
Phase~A, the primary ZIP detector development 
requirement for SuperCDMS is to increase the detector thickness from the present 10~mm to 
25~mm. This will both give an effective increase of a factor 2.5 in 
detector rejection capability for beta-like surface events, and increase 
the throughput in detector fabrication, testing, and calibration.

The remaining factor of two in detector background rejection capability we 
believe will come from optimizing 
the hydrogenation of the amorphous-Si electrodes used for the ionization signal 
measurement. For technical reasons hydrogenation of the amorphous Si for 
the ZIP detectors deployed in CDMS~II at Soudan was not possible. However, 
earlier detectors used by CDMS had higher ionization-only based rejection 
capability, thus our interest in developing this option further. 

Further improvements in software analysis of the athermal phonon pulse shapes will also 
increase our background rejection ability. Most of the desired improvement 
(factor of four) for SuperCDMS Phase~A (relative to the first analysis of 
CDMS~II Soudan data~\cite{PRL2004}) is already close at hand.        

Finally, we are also investigating the possible sources of beta 
contamination present in the first CDMS~II detectors run at Soudan. The 
dominant radioactive background appears to be Pb-210, most likely from Radon 
plate-out. Studies\cite{PRD2005} of the Soudan data, Monte-Carlo simulations of possible sources, 
and surface-analysis of test samples appear to indicate that the other likely candidates,
K-40  and C-14, are not significant ($< 20\%$ of the present total) so no 
remedial actions are required for those sources prior to SuperCDMS Phase~A. 
During the course of CDMS~II more rigorous procedures were put in place to 
mitigate radon plate-out. The impact of these efforts will only be known as 
later detector towers of CDMS~II come into operation at Soudan over the 
next year. Of course it is possible that a significant fraction of the 
present beta contamination is due to a source not yet identified. 
Inductively-coupled plasma mass spectroscopy (ICP-MS) is a quick method for 
identifying most of the potential exotics. Alpha, beta and gamma 
screeners are able to identify the others.
     
Note that it is only the combined affect of 
these three approaches to the misidentified beta-contamination background that matters:
some approaches will be more or less successful than originally 
anticipated,
but they will all be pursued under the SuperCDMS 
program to ensure that the overall risk is minimized.

\subsection{SuperCDMS Development Project}

The first part of SuperCDMS, proposed jointly to the DOE \& NSF in late 2004,
is the Development Project that will demonstrate that 
all the detector advances required for Phase~A (outlined above) have been 
met.

An additional goal of the Development Project indicated in 
Fig.~\ref{fig:timeline} is to continue our present operations at Soudan 
to extend our WIMP-search reach beyond the CDMS~II goals. 
The first new ZIP-style Ge detectors of SuperCDMS would also first be tested at Soudan 
(designated ``ST$_6~1$'' in Fig.~\ref{fig:timeline}), in the present Soudan 
ice-box along-with the incumbent 
five towers of CDMS~II detectors (designated as T* in Fig.~\ref{fig:timeline})
and thus allow further science returns prior to a deeper site, such as 
SNOLab, becoming operational.

The SuperCDMS Development Project would
also allow development of additional longer lead-time items that would be 
required to achieve the SuperCDMS Phases B \& C goals. These include
both the increased zero-background goals, and accelerated detector 
manufacturing rates. 

\subsection{SuperCDMS Apparatus Concepts}

Here we will outline the longer-term infrastructure needs of SuperCDMS and 
highlight some of the elements we consider necessary for the later Phases.

\subsubsection{Cryogenics}

\begin{figure}[tbhp]
\centering
\includegraphics[width=62mm]{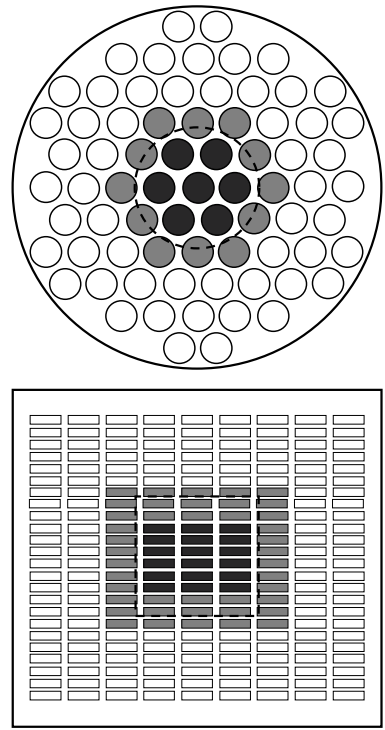}
\caption{Plan and elevation views of the SuperCDMS cold-volume showing 
deployment of 7 towers of 6 detectors each for SuperCDMS Phase~A (darkest circles), 19 towers 
of 12 detectors each in the 145~kg Phase~B (grey circles), and 73 towers 
of 24 detectors each in the 1.1 tonne Phase~C (open circles). The 
required cold-volume will be three times larger in each dimension (dashes) 
than that of CDMS.
} \label{fig:phases}
\end{figure}

In order to deploy the larger mass of Ge detectors for SuperCDMS 
a new, larger cryogenic facility at a new, deeper site is required. To 
accommodate 1000~kg-worth of Ge detectors requires a cold volume of 
$\simeq 1\,{\rm m}^3$. Figure~\ref{fig:phases} shows a possible 
arrangement of the 25~mm thick ZIP-style detectors of SuperCDMS for each 
phase of the program within the new cold volume.

\begin{figure*}[htbp]
\centering
\includegraphics[width=140mm]{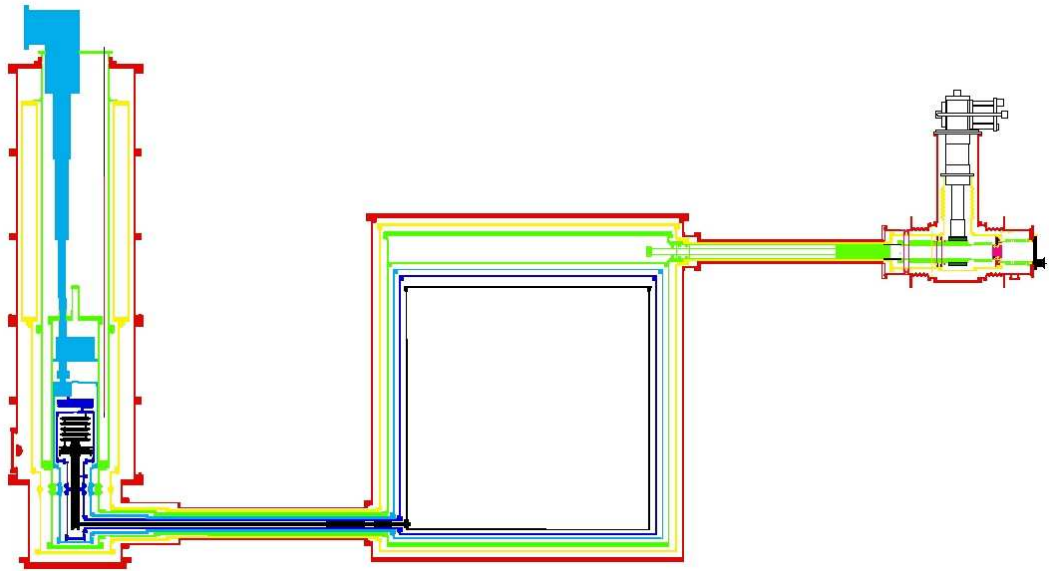}
\caption{Design concept for cryogenic facility showing (left to right) 
a commercial dilution refrigerator connected to the radio-pure copper vessel 
containing the Ge detectors, with electrical readout through the stem to the
right cooled by a closed-cycle cryocooler.
} \label{fig:SNOBox}
\end{figure*}

A conceptual sketch of the envisioned overall cryogenic 
design is shown in Fig.~\ref{fig:SNOBox} and is similar to the present 
CDMS~II arrangement at Soudan~\cite{Ogburn}. 
An MRI 
proposal to the NSF for both the design and 
construction of the desired system was submitted in early 2005. One of the 
design philosophies is that the mounting and electrical readout of 
detectors is modular, so that some flexibility is possible in the 
utilization of the cryogenic facility.

\subsubsection{Electronics}

The readout of the ionization signals in CDMS presently requires FETs 
self-heated to 140~K relatively close to the detectors. As we 
increase the number of detectors for the later phases of SuperCDMS, the 
heat-load  will 
necessitate the replacement of the FET readout with the lower power 
dissipation of SQUID-based charge readout~\cite{QSQUID}. Such concepts require 
further development under the SuperCDMS program and related long-term
R~\&~D proposals. 

As the number of phonon-signal readout channels is increased there is 
additional motivation to move from the present CDMS single-stage SQUID 
readout to two-stage configurations. Apart from ease in manufacture and
demonstrated multiplexing schemes~\cite{two-stage}, 
opportunities are present for improving the detector phonon sensor design. 
Larger phonon collection area designs would be possible and 
substitution of the present W Transition-Edge-Sensor (TES) elements with 
Al-Mn TESs~\cite{Al-Mn} would also be possible. These avenues are of 
interest both for improving the phonon sensor performance and easing the 
manufacture of future ZIP-style detectors.

\subsubsection{Detectors}

An alternative configuration of the ionization and phonon channels to that 
of CDMS~II had been suggested by Luke~\cite{Luke} where the ionization bias 
electrodes are interdigitated with the phonon sensors such that electric 
fields are now present tangentially to the detector surface in addition to the 
usual bulk drift field through the detector. This would allow vetoing of 
ionizing surface events. We suspect that such a scheme may be required to 
achieve the detector performance goals of SuperCDMS Phase~B.

Alternative double-sided phonon sensor readout schemes would also improve our 
utilization of athermal phonon signal pulse-shape analysis and event 
reconstruction. The biasing of the CDMS~II-style ionization electrodes would 
be more complicated for these schemes but is being studied under long-term 
R~\&~D base programs.

\section{Conclusion}

The potential of the CDMS ZIP detectors has become readily apparent 
over the last few years, with CDMS~II presently the world-leading 
WIMP-search direction-detection experiment~\cite{PRL2004}.
The measurement of {\it both} ionization and phonon signals for each 
event avoids possible ambiguities at the low recoil of 
energies of interest. A positive signal identification from {\it any} 
WIMP-search experiment must be compelling - with no other alternative 
sources of events, detector pathologies, or systematic affects clouding the issue. We believe 
SuperCDMS will satisfy these demands. 

SuperCDMS is by no means the only way forward, but we believe that it will
be a strong contender with other 
large-mass detector concepts~\cite{Others}. 

Indeed, comparison of any WIMP discovery by complementary detector technologies is 
a scientific necessity and will also aid in refining properties of the 
WIMPs, for example their mass, velocity distribution and coupling 
strengths to nucleonic matter. The age of WIMP astronomy is approaching.



\bigskip 
\begin{acknowledgments}

This work is supported by the National Science 
Foundation (NSF) under Grant No. AST-9978911, by the 
Department of Energy under contracts DE-AC03-76SF00098,
DE-FG03-90ER40569, DE-FG03-91ER40618, and by
Fermilab, operated by the Universities Research
Association, Inc., under Contract No. DE-AC02-76CH03000
with the Department of Energy. 
The ZIP detectors are fabricated in the Stanford Nanofabrication Facility (which 
is a member of the National Nanofabrication Infrastructure Network sponsored 
by NSF under Grant ECS-0335765).     
In addition, seed funding for
SuperCDMS detector development has been provided 
at Stanford by the KIPAC Enterprise Fund, the Dean
of Research, and a
Center for Integrated Systems Seed Grant.

\end{acknowledgments}

\bigskip 

\end{document}